\documentclass[aps,prb,groupedaddress,showkeys,showpacs,twocolumn,floatfix]{revtex4}% Physical Review B

\usepackage{amssymb,amsbsy,amsmath}

\usepackage{graphicx}% Include figure files
\usepackage{dcolumn}% Align table columns on decimal point
\usepackage{bm}% bold math
\usepackage{subfigure}

\usepackage{multirow}
\usepackage{array}
\usepackage{rotating}

\newcommand{\bra}[1]{\ensuremath{\langle #1|}}
\newcommand{\ket}[1]{\ensuremath{|#1\rangle}}
\newcommand{\braket}[2]{\ensuremath{\langle #1|#2\rangle}}

\newcommand{\op}[1]{%
    \fontdimen12\textfont3=2pt\fontdimen12\scriptfont3=1.4pt%
    \!\null\mathop{\vphantom{#1}\smash{#1}}\limits_{\sim}\null\!}
\newcommand{\vecop}[1]{%
    \fontdimen12\textfont3=2pt\fontdimen12\scriptfont3=1.4pt%
    \!\null\mathop{\textbf{\vphantom{#1}\smash{#1}}}\limits_{\sim}\null\!}

\newcommand{\fmref}[1]{(\protect\ref{#1})}
\newcommand{\xref}[1]{\protect\ref{#1}}
\newcommand{\figref}[1]{Fig.~\protect\ref{#1}}

\newcommand{\reduced}[3]{\ensuremath{\langle #1||#2 || #3 \rangle}}

\newcommand{\com}[2]{\ensuremath{\left[ #1,\,#2 \right]}}

\newcommand{\threej}[6]{\ensuremath{\begin{pmatrix} &#1& \quad &#2& \quad &#3& \\ &#4& \quad &#5& \quad &#6& \end{pmatrix}}}

\newcommand{\ito}[3]{\ensuremath{\op{#1}^{(#2)}_{#3}}}
\newcommand{\vecito}[2]{\ensuremath{\vecop{#1}^{(#2)}}}

\begin{document}

\title{Numerically exact and approximate determination of
  energy eigenvalues for antiferromagnetic molecules using
  irreducible tensor operators and general point-group
  symmetries}% Force line breaks with \\

\author{Roman Schnalle}
\email{rschnall@uos.de}
\affiliation{Universit{\"a}t Osnabr{\"u}ck, Fachbereich Physik, D-49069 Osnabr{\"u}ck, Germany}
\author{J{\"u}rgen Schnack}
\email{jschnack@physik.uni-bielefeld.de}
\affiliation{Universit{\"a}t Bielefeld, Fakult{\"a}t f{\"u}r Physik, Postfach 100131, D-33501 Bielefeld, Germany}

\date{\today}

\begin{abstract}
Numerical exact diagonalization is the ultimate method of choice
in order to discuss static, dynamic, and thermodynamic
properties of quantum systems. In this article we consider
Heisenberg spin-systems and extend the range of applicability of
the exact diagonalization method by showing how the irreducible
tensor operator technique can be combined with an unrestricted
use of general point-group symmetries. We also present ideas how
to use spin-rotational and point-group symmetries in order to
obtain approximate spectra.
\end{abstract}


\maketitle

%%%%%%%%%%%%%%%%%%%%%%%%%%%%%%%%%%%%%%%%%%%%%%%%%%%%%%%%%%%%%%%%%%%%%%%%
\section{Introduction}
\label{sec-1}

The knowledge of energy spectra of small magnetic systems such
as magnetic molecules is indispensable for the (complete)
understanding of their spectroscopic, dynamic, and thermodynamic
properties. In this respect numerical exact diagonalization of
the appropriate quantum Hamiltonian is the ultimate method of
choice. Nevertheless, such an attempt is very often severely
restricted due to the huge dimension of the underlying Hilbert
space. For a magnetic system of $N$ spins of spin quantum number
$s$ the dimension is $(2s+1)^N$ which grows exponentially with
$N$. Group theoretical methods can help to ease this numerical
problem. A further benefit is given by the characterization of
the obtained energy levels by quantum numbers and classification
according to irreducible representations.

Along these lines much effort has been put into the development
of an efficient numerical diagonalization technique of the
Heisenberg model using irreducible tensor operators,
i.e. employing the full rotational symmetry of angular
momenta.\cite{GaP:GCI93,BCC:IC99,BeG:EPR,Tsu:group_theory,Tsu:ICA08}
A combination of this meanwhile well established technique with
point-group symmetries is not very common since a rearrangement
of spins due to point-group operations easily leads to
complicated basis transformations between different coupling
schemes. A possible compromise is to use only part of the
spin-rotational symmetry (namely rotations about the $z$--axis)
together with point-group symmetries\cite{RLM:PRB08} or to
expand all basis states in terms of simpler product
states.\cite{PhysRevB.64.064419,PhysRevB.68.029902,PhysRevB.64.014408}
To the best of our knowledge only two groups developed a
procedure in which the full spin-rotational symmetry is combined
with point-group symmetries. O.~Waldmann combines the full
spin-rotational symmetry with those point-group symmetries that
are compatible with the spin coupling scheme, i.e. avoid
complicated basis transforms between different coupling
schemes.\cite{Wal:PRB00} Sinitsyn, Bostrem, and Ovchinnikov
follow a similar route for the square lattice antiferromagnet by
employing $D_4$ point-group symmetry.\cite{BOS:TMP06,SBO:JPA07}
This already establishes a very powerful numerical method.

In this article we show how the irreducible tensor operator
technique can be combined with an unrestricted use of general
point-group symmetries. The problem, that the application of
point-group operations leads to states belonging to a basis
characterized by a different coupling scheme whose representation
in the original basis is not (easily) known, can be solved by
means of graph theoretical methods that have been developed in
another context.\cite{FPV:CPC97,FPV:CPC95} We discuss how this
methods can be implemented and present results for numerical
exact diagonalizations of Heisenberg spin systems of
unprecedented size.

Having these methods developed we also discuss ideas of
approximately obtaining energy spectra of so-called bipartite,
i.e. non-frustrated, antiferromagnetic spin systems. The idea is
to perform numerical diagonalizations in the orthogonal Hilbert
subspaces characterized by spin and point-group quantum numbers
using only a restricted but carefully chosen basis subset. We
demonstrate how this idea works for archetypical spin systems
such as bipartite or slightly frustrated spin rings. The
advantage compared to alternative approximate methods such as
Density Matrix Renormalization
Group\cite{Whi:PRB93,ExS:PRB03,Sch:RMP05} (DMRG),
Lanczos,\cite{Lan:JRNBS50} or Quantum Monte
Carlo\cite{SaK:PRB91,San:PRB99,EnL:PRB06} (QMC) techniques is,
that one obtains many energy levels together with their
spectroscopic classification which can be of great use for the
discussion of Electron Paramagnetic Resonance (EPR), Nuclear
Magnetic Resonance (NMR), or Inelastic Neutron Scattering (INS)
spectra. In this respect our idea can provide a valuable
complement to the already established approximate methods.

The article is organized as follows. In Sec.~\xref{sec-2} we
explain the idea of a combined usage of spin-rotational and
point-group symmetry. Section~\xref{sec-3} provides examples
for full diagonalization studies. Our approximate
diagonalization scheme is introduced in Sec.~\xref{sec-4},
whereas Sec.~\xref{sec-5} provides example calculations on
bipartite systems. The paper closes with a summary.

%%%%%%%%%%%%%%%%%%%%%%%%%%%%%%%%%%%%%%%%%%%%%%%%%%%%%%%%%%%%%%%%%%%%%%%%
\section{Theoretical method}
\label{sec-2}

%%%%%%%%%%%%%%%%%%%%%%%%%%%%%%%%%%%%%%%%%%%%%%%%%%%%%%%%%%%%%%%%%%%%%%%%
\subsection{Irreducible tensor operator approach}

The physics of many magnetic molecules can be well understood
with the help of the isotropic Heisenberg model with
nearest-neighbor coupling. The action of an external magnetic
field is accounted for by an additional Zeeman term. The
resulting Hamiltonian then looks like
%--------------------------------------------------------
\begin{equation} \label{eq:Hamiltonoperator}
   \op{H}= - \sum_{i,j} J_{ij} \vecop{s}(i) \cdot \vecop{s}(j) +
   g \mu_B \vecop{S} \cdot \vec{B}
\ .
\end{equation}
%--------------------------------------------------------
The sum reflects the exchange interaction between single spins
given by spin operators $\vecop{s}$ at sites $i$ and $j$.  For
the sake of simplicity we assume a common isotropic $g$--tensor.
Then the Zeeman term couples the total spin operator
$\vecop{S}=\sum_{i=1}^N \vecop{s}(i)$ to the external magnetic
field $\vec{B}$. A negative value of $J_{ij}$ refers to an
antiferromagnetic coupling.

For the following discussion an antiferromagnetic
nearest-neighbor exchange coupling of constant value $J<0$ is
assumed (which can easily be generalized), then the Heisenberg
part can be written as
%--------------------------------------------------------
\begin{equation} \label{eq:Heisenberg_WW}
   \op{H}_\text{Heisenberg} = -J \sum_{<i,j>} \vecop{s}(i) \cdot
   \vecop{s}(j)
\ ,
\end{equation}
%--------------------------------------------------------
where the summation parameter $<i,j>$ indicates the summation
running over nearest-neighbor spins counting each pair only
once. Since the commutation relations
%--------------------------------------------------------
\begin{equation} \label{eq:Kommutatior_Gesamtspin}
\com{\op{H}_\text{Heisenberg}}{\vecop{S}} = 0
\end{equation}
%--------------------------------------------------------
hold it is possible to find a common eigenbasis $\{ \ket{\nu}
\}$ of $\op{H}_\text{Heisenberg}$, $\vecop{S}^2$ and
$\op{S}_z$. We denote the corresponding eigenvalues as $E_\nu$,
$S_\nu$ and $M_\nu$.  Due to spin-rotational symmetry,
Eq. \fmref{eq:Kommutatior_Gesamtspin}, the eigenvalues of the
Hamiltonian \fmref{eq:Hamiltonoperator} can be evaluated (later)
according to
%--------------------------------------------------------
\begin{equation}
   E_\nu(B)=E_\nu + g \mu_B B M_\nu
\ ,
\end{equation}
%--------------------------------------------------------
where the direction of the external field $\vec{B}$ defines the
 $z$-axis.

Calculating the eigenvalues here corresponds to finding a matrix
representation of the Hamiltonian and diagonalizing it
numerically. A very efficient and elegant way of finding the
matrix elements of Eq. \fmref{eq:Heisenberg_WW} is based on the
use of irreducible tensor operators. Apart from its elegance it
drastically reduces the dimensionality of the problem because it
becomes possible to work directly within the subspace
$\mathcal{H}(S,M=S)$ of the total Hilbert space $\mathcal{H}$
characterized by quantum numbers $S$ and $M=S$; for typical
dimensions compare for instance Ref.~\onlinecite{BSS:JMMM00}.

The calculation of matrix elements of
the given Hamiltonian using irreducible tensor operators is
compulsorily related to the application of the Wigner-Eckart-theorem. The Wigner-Eckart-theorem
%--------------------------------------------------------
\begin{eqnarray} \label{eq:wigner-eckart}
  && \bra{\alpha \, S \, M} \ito{T}{k}{q} \ket{\alpha' \, S' \, M'} = \nonumber \\
  && (-1)^{S-M} \reduced{\alpha \, S}{\vecito{T}{k}}{\alpha' \, S'} \threej{S}{k}{S'}{-M}{q}{M'}
\end{eqnarray}
%--------------------------------------------------------
states that a matrix element of the $q$-th component of an
irreducible tensor operator $\vecito{T}{k}$ of rank $k$ is given
by the reduced matrix element $\reduced{\alpha \,
S}{\vecito{T}{k}}{\alpha' \, S'}$ and a factor containing a
Wigner-3J symbol.\cite{VMK:quantum_theory}

It should be emphasized that the reduced matrix element is
completely independent of any magnetic quantum number $M$. The
basis in Eq. \fmref{eq:wigner-eckart} is given following the
well-known vector-coupling-scheme. The quantum number $\alpha$
within the ket $\ket{\alpha \, S \, M}$ refers to a set of
intermediate spin quantum numbers resulting from the coupling of
single spins $s$ to the total spin quantum number $S$. In order
to apply the Wigner-Eckart-theorem it is necessary to express
the Heisenberg Hamiltonian in Eq. \fmref{eq:Heisenberg_WW} with
the help of irreducible tensor operators. Therefore the
single-spin vector operators $\vecop{s}(i)$ can be seen as
irreducible tensor operators of rank $k=1$ with components
$q=-1,0,1$. The relation to the components of the vector
operators is given by
%--------------------------------------------------------
\begin{equation} \label{eq:spintensors}
   \ito{s}{1}{0} = \op{s}^z,\quad \ito{s}{1}{\pm 1} = \mp
   \sqrt{\frac{1}{2}} \left( \op{s}^x \pm i \op{s}^y \right) 
\ .
\end{equation}
%--------------------------------------------------------
Writing the Heisenberg exchange term as a tensor product of the
single-spin irreducible tensor operators results
in\cite{BCC:IC99}
%--------------------------------------------------------
\begin{eqnarray} \label{eq:Tensor_Heisenberg}
   && \op{H}_\text{Heisenberg} = \nonumber \\
   && \quad \sqrt{3} J \sum_{<i,j>} \ito{T}{0}{}(\{ k_l \},\{
   \overline{k}_m \}|k_i=k_j=1)
\ .
\end{eqnarray}
%--------------------------------------------------------
$\ito{T}{0}{}$ is a zero-rank irreducible tensor operator
depending on the sets $\{ k_l \}$, $l=1,\dots,N$, which give
the ranks of single-spin irreducible tensor operators and $\{
\overline{k}_m \}$, $m=1,\dots,N-1$, which refers to the ranks
of intermediate irreducible tensor operators. In a successive
coupling scheme within a system of $N$ spins an irreducible
tensor operator of this kind would look like
%--------------------------------------------------------
\begin{eqnarray} \label{eq:allg_coupl}
   && \ito{T}{0}{} (\{ k_l \},\{ \overline{k}_m \}) =  
\{ \dots \{ \{ \vecop{s}^{(k_1)}(1) \otimes \vecop{s}^{(k_2)}(2) \}^{(\overline{k}_{1})} \otimes \nonumber \\
   && \quad \vecop{s}^{(k_3)}(3) \}^{(\overline{k}_2)} \dots
\}^{(\overline{k}_{N-2})} \otimes \vecop{s}^{(k_N)}(N) \}^{(0)}
\ .
\end{eqnarray}
%--------------------------------------------------------
The notation $\ito{T}{0}{}(\{ k_l \},\{ \overline{k}_m
\}|k_i=k_j=1)$ corresponds to the situation in which the ranks
of all single-spin tensor operators are zero except those at
sites $i$ and $j$ which are tensor operators of rank $1$.

The set $\{ \overline{k}_m \}$ results from the chosen coupling
scheme, for example of the form of Eq. \fmref{eq:allg_coupl},
with known ranks of single-spin tensor operators taking into
account addition rules for spin quantum numbers of the vector
coupling scheme like $\overline{k}_1=|k_1-k_2|,\dots,k_1 +
k_2$.

After writing the Heisenberg Hamiltonian as a sum of irreducible
tensor operators the matrix elements within a basis of the form
$\ket{\alpha \, S \, M}$ can be calculated by the application of
the Wigner-Eckart-theorem. The reduced matrix elements are
determined using the so-called \textit{decoupling}
procedure.\cite{BeG:EPR,Tsu:group_theory} Since the irreducible
tensor operator $\vecito{T}{k}$ is given as a tensor product of
irreducible tensor operators with regard to a certain coupling
scheme (comp. Eq. \fmref{eq:allg_coupl}), the reduced matrix
element $\reduced{\alpha \, S}{\vecito{T}{k}}{\alpha' \, S'}$
can successively be decomposed into a product of single-spin
irreducible tensor operators and Wigner-9J symbols.

%%%%%%%%%%%%%%%%%%%%%%%%%%%%%%%%%%%%%%%%%%%%%%%%%%%%%%%%%%%%%%%%%%%%%%%%
\subsection{General point-group symmetries}

The use of irreducible tensor operators for the calculation of
the matrix elements of the Hamiltonian and as a result also of
the energy spectrum is essential for the treatment of magnetic
molecules containing many interacting paramagnetic
ions. Nevertheless, it is sometimes necessary to further reduce
the dimensionality of the problem, either because computational
resources are limited or a labeling of certain energy levels
becomes advantageous, e.g. for spectroscopic
classification.\cite{Gri:SB72,TBF:SSRBC87} Such a reduction can
be done if the Hamiltonian remains invariant under certain
permutations of spin centers. Often the spin-permutational
symmetry of the Hamiltonian coincides with spatial symmetries of
the molecule, i.e. point-group symmetries, therefore the term
point-group symmetry is used while one refers to the invariance
of the Hamiltonian under permutations of spins.

Using point-group symmetries of the system results in a
decomposition of the Hamilton matrix $\bra{\alpha \, S \,
M}\op{H}\ket{\alpha' \, S \, M}$ into irreducible
representations $\Gamma^{(n)}(\mathcal{G})$ of a group
$\mathcal{G}$ whose elements $\op{G}(R)$, i.e. the operators
corresponding to the symmetry operations $R$, do commute with
$\op{H}$.

The symmetrized basis functions which span the irreducible
representations $n$ are found by the application of the
projection operator $\mathcal{P}^{(n)}$ to the full set of basis
vectors $\ket{\alpha \, S \, M}$ and subsequent
orthonormalization. The over-complete set of basis states $\{
\ket{\alpha \, S \, M \, \Gamma^{(n)}} \}$ spanning the $n$-th
irreducible representation $\Gamma^{(n)}(\mathcal{G})$ is
generated by\cite{Tin:Group_theory}
%--------------------------------------------------------
\begin{equation} \label{eq:Projektionsoperator}
   \mathcal{P}^{(n)}\ket{\alpha \, S \, M} = 
\left( \frac{l_n}{h} \sum_{R} \left(\chi^{(n)}(R)\right)^\ast 
\, \op{G}(R) \right) \ket{\alpha \, S \, M}
\ ,
\end{equation}
%--------------------------------------------------------
where $l_n$ is the dimension of the irreducible representation
$\Gamma^{(n)}$, $h$ denotes the order of $\mathcal{G}$ and
$\chi^{(n)}(R)$ is the character of the $n$-th irreducible
representation of the symmetry operation $R$.

Equation \fmref{eq:Projektionsoperator} contains the main challenge
while creating symmetrized basis states. The action of the
operators $\op{G}(R)$ on basis states of the form $\ket{\alpha
\, S \, M}$ has to be known. Of course, one could expand
$\ket{\alpha \, S \, M}$ into a linear combination of product
states $\ket{m_1 \, m_2 \dots m_N}$. Then the action of
$\op{G}(R)$ leads to a permutation of magnetic quantum numbers
$m_i$ within the ket $\ket{m_1 \, m_2 \dots m_N}$. But, the
recombination of the symmetry-transformed product states into
basis states $\ket{\alpha \, S \, M}$ will then be much too time
consuming for larger systems.

Following Ref. \onlinecite{Wal:PRB00} the action of $\op{G}(R)$
on states $\ket{\alpha \, S \, M}$ can directly be evaluated
without expanding it into product states. Suppose there is a
certain coupling scheme $a$ in which spin operators
$\vecop{s}(i)$ are coupled to yield the total spin operator
$\vecop{S}$. Generally the action of operators $\op{G}(R)$ on
states $\ket{\alpha \, S \, M}$ leads to a different coupling
scheme $b$. Now those states which belong to the coupling scheme
$b$ have to be reconverted into a linear combination of states
belonging to $a$. This is technically a rather involved
calculation, and one would not like to do it by hand and for
every coupling scheme separately. To the best of our knowledge
it has never been noted or even used that the conversion from
any arbitrary (!) coupling scheme $b$ into the desired coupling
scheme $a$ can be well automatized. Suppose there is a state
$\ket{\alpha \, S \, M}_a$ belonging to the coupling scheme
$a$. The action of a arbitrary group element $\op{G}(R)$ results
in a state $\ket{\alpha \, S \, M}_b$ belonging to a different
coupling scheme $b$. Then the re-expression takes the following
form
%--------------------------------------------------------
\begin{eqnarray} \label{eq:gen_rec_coeff}
   &&\op{G}(R) \ket{\alpha \, S \, M}_a = \nonumber \\
   && \quad \sum_{\alpha'} \ket{\alpha' \, S \, M}_a \;
   _a\braket{\alpha' \, S \, M}{\alpha \, S \, M}_b
\ ,
\end{eqnarray}
%--------------------------------------------------------
where a term like $_a\braket{\alpha' \, S \, M}{\beta \, S \,
M}_b$ is known as \textit{general re-coupling coefficient}. The
calculation of general re-coupling coefficients and the
evaluation of Eq. \fmref{eq:gen_rec_coeff} can be performed with
the help of graph theoretical methods.\cite{FPV:CPC97,FPV:CPC95}
An implementation of these methods within a computer program is
a straightforward task (follow directions given in
Refs.~\onlinecite{FPV:CPC97,FPV:CPC95}) and one can deal with
any point-group symmetry.

%%%%%%%%%%%%%%%%%%%%%%%%%%%%%%%%%%%%%%%%%%%%%%%%%%%%%%%%%%%%%%%%%%%%%%%%
\section{Numerical exact diagonalization}
\label{sec-3}

%===================    figure   =================================
\begin{figure}[ht!]
\centering
\includegraphics[clip,width=32mm]{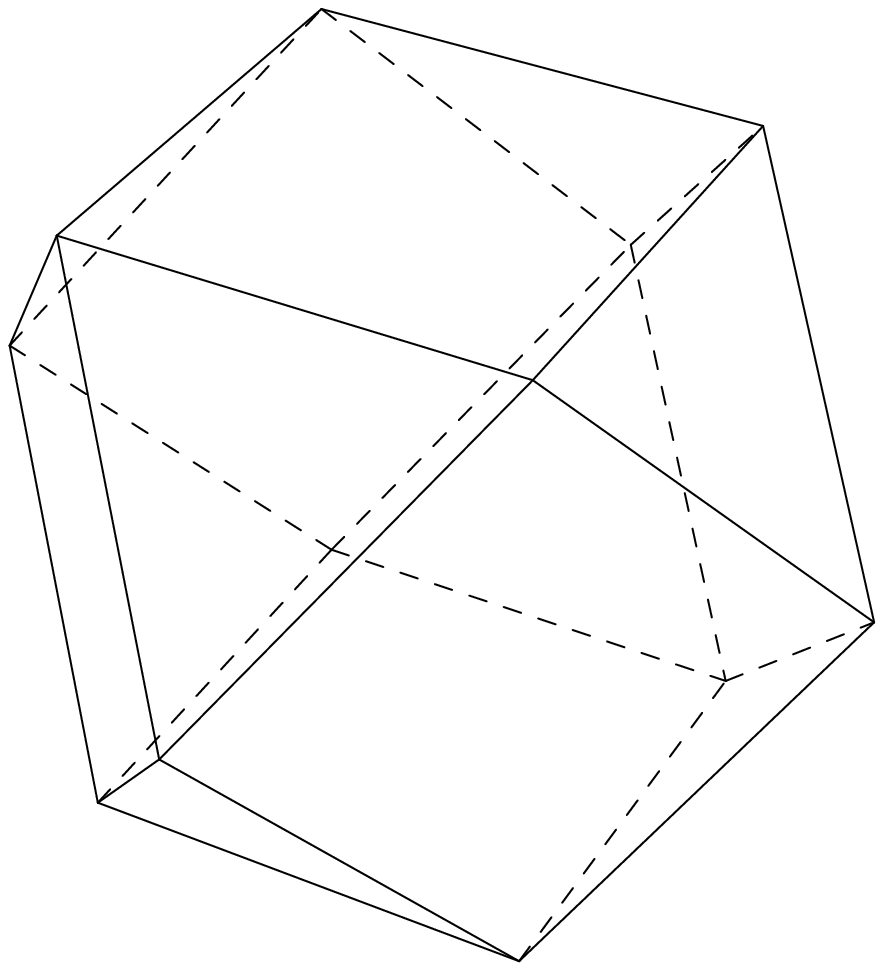}
\quad
\includegraphics[clip,width=35mm]{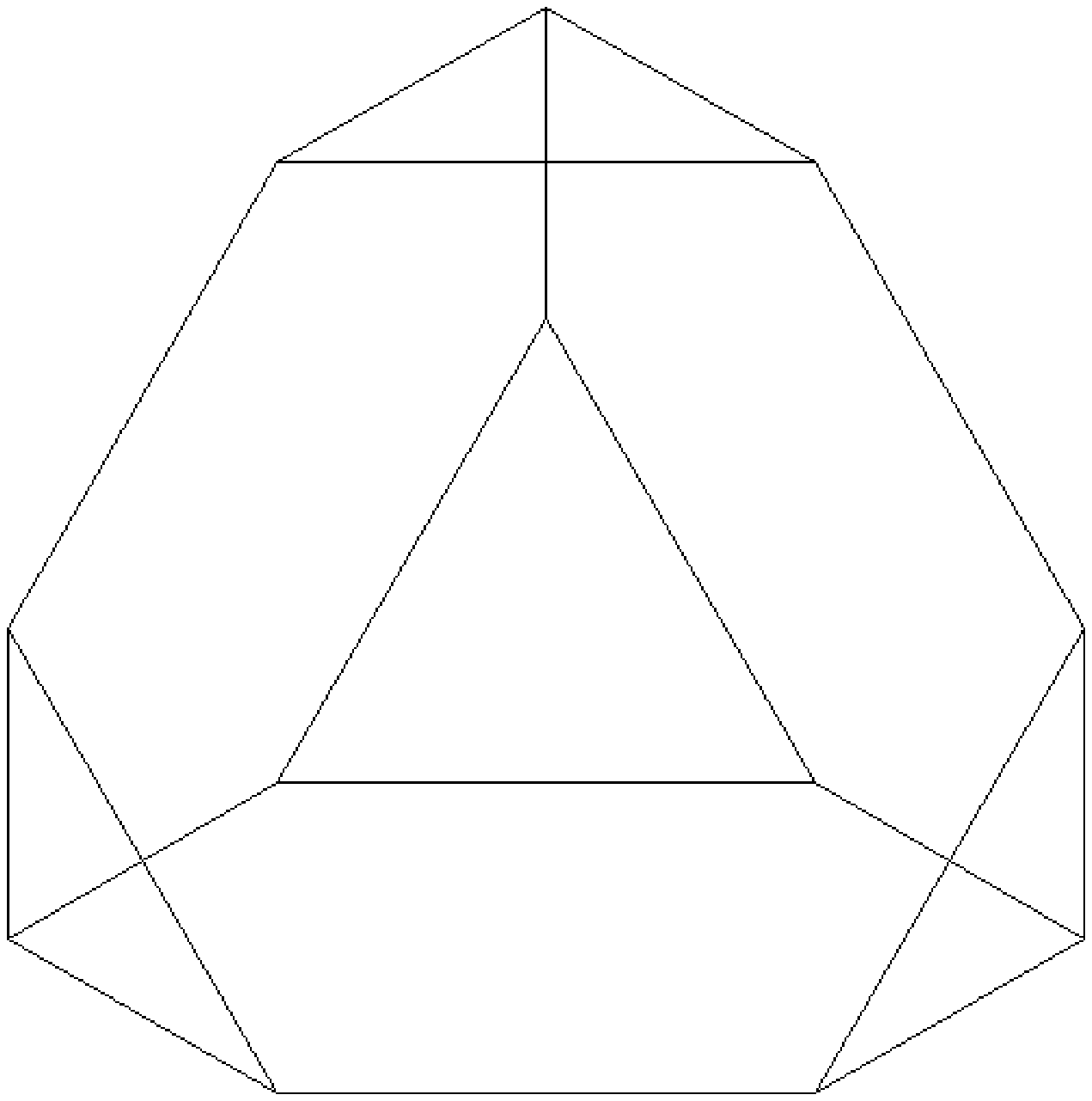}
\caption{Structure of the cuboctahedron (l.h.s.) and the
  truncated tetrahedron (r.h.s.).\cite{mathworld}}
\label{F-1}
\end{figure}
%===================    figure  =================================

In this section we like to present two applications for
realistic spin systems that can be treated using irreducible
tensor operator techniques and point-group symmetries, but not
otherwise. Both systems -- cuboctahedron and truncated
tetrahedron -- consist of $N=12$ spins of spin quantum
number $s=3/2$ (Hilbert space dimension 16,777,216). The
two spin systems, which are realized as antiferromagnetic
molecules -- cuboctahedron\cite{BGG:JCSDT97} and truncated
tetrahedron\cite{PLK:CC07}, see \figref{F-1} for the structure
-- belong to the class of geometrically frustrated spin
systems\cite{SSR:JMMM05,SSR:PRB07,ScS:P08} and are thus hardly
accessible by means of Quantum Monte Carlo.

%===================    figure   =================================
\begin{figure}[ht]
   \centering
   \includegraphics[width=7.5cm]{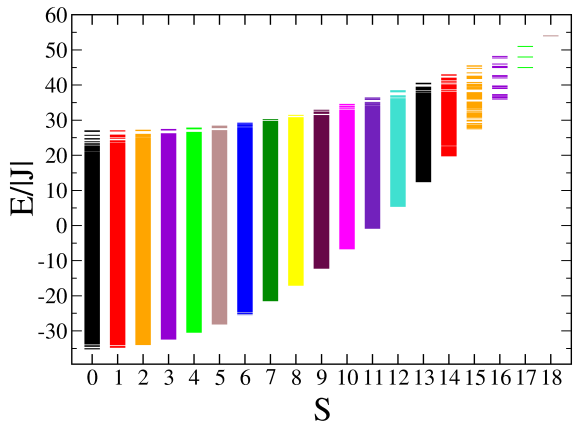}
   \caption{(Color online) Complete energy spectrum of the antiferromagnetic
   cuboctahedron with $s=3/2$.}
   \label{F-2}
\end{figure}
%===================    figure   =================================

\begin{figure}[ht]
   \centering
   \includegraphics[width=7.5cm]{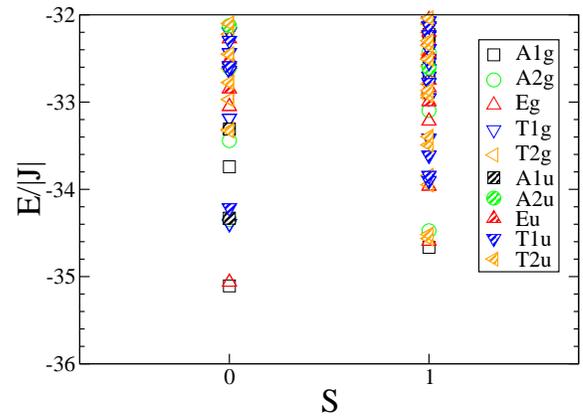}
   \caption{(Color online) Low-lying energy spectrum of the
   antiferromagnetic 
   cuboctahedron with $s=3/2$ in subspaces of $S=0$ and $S=1$. The
   symbols denote the irreducible representations of the $O_h$
   group.} 
   \label{F-3}
\end{figure}

Figure~\xref{F-2} shows the energy spectrum of the
antiferromagnetic cuboctahedron with $s=3/2$. This spectrum was
obtained using only $D_2$ point-group symmetry which is already sufficient
in order to obtain sufficiently small Hamilton matrices. In
addition \figref{F-3} demonstrates for the
subspaces of total spin $S=0$ and $S=1$ that a representation in
the full $O_h$ group can be achieved which yields level
assignments according to the irreducible representations of this
group.

%===================    figure   =================================
%===================    figure   =================================

\begin{figure}[ht]
   \centering
   \includegraphics[width=7.5cm]{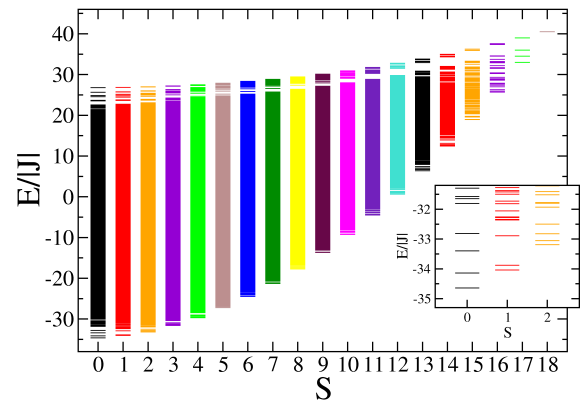}
   \caption{(Color online) Complete energy spectrum of the
   antiferromagnetic truncated tetrahedron with $J_1=J_2=J$. The
   inset shows low-lying levels in subspaces with $S=0, 1, 2$.}
   \label{F-4}
\end{figure}

A complete energy spectrum allows to calculate thermodynamic
properties as functions of both temperature $T$ and magnetic
field $B$. For the cuboctahedron this was already done
elsewhere.\cite{ScS:P08} Therefore, we like to discuss another
frustrated structure, the truncated tetrahedron which was
synthesized quite recently.\cite{PLK:CC07} In principle this
geometry permits two different exchange constants, one inside
the triangles ($J_1$) and the other between the triangles
($J_2$), compare \figref{F-1}. A practical symmetry for this
molecule is for instance $C_{2v}$, whereas the full symmetry is
$T_d$.  Figure~\xref{F-4} displays the complete
energy spectrum for the case $J_1=J_2=J$. The inset of
\figref{F-4} magnifies the low-energy sector. As in
the case of many other frustrated antiferromagnetic systems the
spectrum exhibits more than one singlet below the first
triplet.\cite{SSR:JMMM05}

%===================    figure   =================================
%===================    figure   =================================

\begin{figure}[ht]
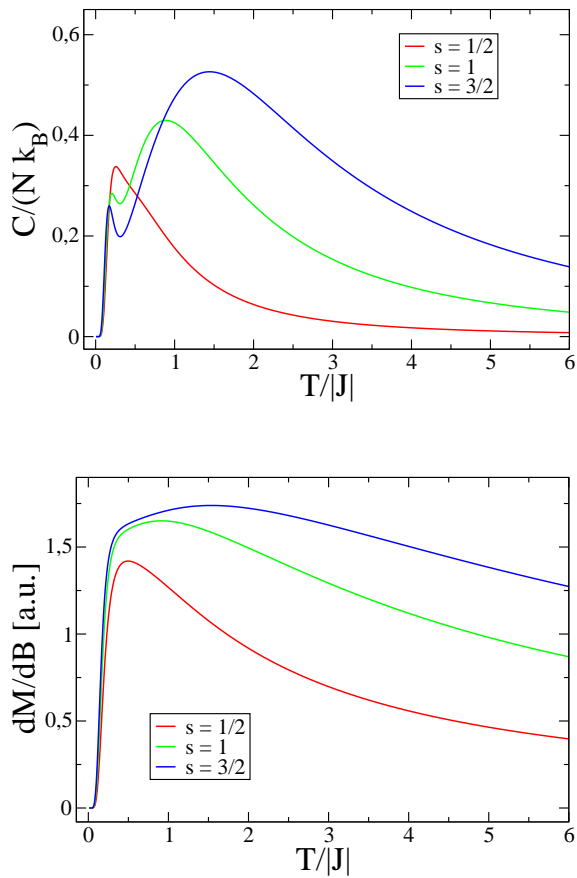

   \centering
   \includegraphics[width=7.5cm]{ito-fig-5a.eps}

   \vspace*{10mm}

   \includegraphics[width=7.5cm]{ito-fig-5b.eps}
   \caption{(Color online) Specific heat $c(T,B=0)$ (top) and differential
   magnetic susceptibility $\chi(T,B=0)$ (bottom) of the
   truncated tetrahedron with $J_1=J_2=J$.}
   \label{F-5}
\end{figure}

In \figref{F-5} we show the zero-field specific
heat (top) as well as the zero-field differential magnetic
susceptibility (bottom). The fine structure of the specific
heat, which is especially pronounced for $s=3/2$, results from
the low-energy gap structure. The sharp peak is an outcome of
the gap between the lowest singlet and the group of levels
around the second singlet and the first two triplets, the latter
being highly degenerate (both nine-fold including
$M$-degeneracy). This unusual degeneracy of the lowest triplets
is also the origin of the quick rise and subsequent flat
behavior of the susceptibility in the case of $s=3/2$.

%===================    figure   =================================
%===================    figure   =================================

%%%%%%%%%%%%%%%%%%%%%%%%%%%%%%%%%%%%%%%%%%%%%%%%%%%%%%%%%%%%%%%%%%%%%%%%
\section{Approximate diagonalization}
\label{sec-4}

The previous sections demonstrate that numerical exact
diagonalization in connection with irreducible tensor operators
is a powerful tool to investigate thermodynamical properties of
large magnetic molecules. Nevertheless, sometimes the use of
total-spin and point-group symmetries is not sufficient to
obtain small enough matrices. For such cases we suggest an
approximate diagonalization in this section. The approximation
is partially based on perturbation theory arguments. First ideas
along this line were already suggested in
Ref.~\onlinecite{Wal:PRB07}. We will generalize and largely
extend this idea.

Let's assume that the spin system is described by a Hamiltonian
$\op{H}$ which acts in the Hilbert space $\mathcal{H}$. Suppose
there is a zeroth-order Hamiltonian $\op{H}_0$ and a
decomposition according to
%--------------------------------------------------------
\begin{equation} \label{eq:approxDiag_gen}
   \op{H} = \op{H}_0 + \lambda \op{H}'
\ .
\end{equation}
%--------------------------------------------------------
In the case of non-degenerate eigenstates $\ket{\phi_i^{(0)}}$
of $\op{H}_0$ the series expansion
%--------------------------------------------------------
 \begin{eqnarray} \label{eq:pert_series}
    \ket{\phi_i} &=& \ket{\phi_i^{(0)}} + \lambda
    \ket{\phi_i^{(1)}} 
+ \lambda^2 \ket{\phi_i^{(2)}} + \dots , \label{eq:reihenentwicklung_Phi}\\
    E_i &=& E_i^{(0)} + \lambda E_i^{(1)} + \lambda^2 E_i^{(2)}
    + \dots
\ , \label{eq:reihenentwicklung_E}
 \end{eqnarray}
%--------------------------------------------------------
holds for the exact eigenstates $\ket{\phi_i}$ and corresponding
eigenvalues $E_i$. The index $i=1,\dots,n$ denotes the states of
the system. The energy eigenvalues and eigenstates in
zeroth-order result from a (typically simple or even analytical)
diagonalization of $\op{H}_0$ within an arbitrary basis of
$\mathcal{H}$.

We label the eigenvalues $E_i^{(0)}$ and eigenstates
$\ket{\phi_i^{(0)}}$ in such a manner that
%--------------------------------------------------------
\begin{equation} \label{eq:conv_exp}
   E_i^{(0)} < E_{i+1}^{(0)}, \quad \forall \, i=1,\dots,n-1
\end{equation}
%--------------------------------------------------------
holds. Now we do not follow conventional perturbation theory as
it would lead to a successive introduction of additional terms
within the series expansion in Eq. \fmref{eq:pert_series},
i.e. terms with increasing order of $\lambda$. Instead, we
diagonalize the full Hamiltonian $\op{H}$ within a reduced set
$\{ \ket{\phi_i^{(0)}} \}$, $i=1,\dots,n_\text{red}$, of
eigenstates of $\op{H}_0$, where $n_\text{red} \leq n$ is
referred to as \textit{cut-off parameter}. The resulting
eigenvalues and eigenstates of this approximation are denoted as
$E_i^\text{approx}$ and $\ket{\phi_i^\text{approx}}$. Such an
approximate scheme is always converging since for $n_\text{red}
= n =\text{dim}(\mathcal{H}$) all basis states are incorporated
and the diagonalization corresponds to an exact treatment of the
system, i. e.
%--------------------------------------------------------
\begin{equation}
   E_i^\text{approx} \xrightarrow{n_\text{red} \rightarrow n}
   E_i 
\ , \quad \ket{\phi_i^\text{approx}} \xrightarrow{n_\text{red}
   \rightarrow n} \ket{\phi_i}
\ \forall\ i
\ .
\end{equation}
%--------------------------------------------------------
It is clear that the speed of convergence depends on the choice
of $\op{H}_0$. 

The speed of convergence will be different for the various
states. Since the approximate diagonalization is performed with
the $n_\text{red}$ low-lying states of $\op{H}_0$ according to
\fmref{eq:conv_exp} one expects that the low-lying energy levels
converge quickest against their true values. As in perturbation
theory this assumption relies on the hypothesis that
energetically higher-lying levels do mix into the
desired low-lying state with decreasing weight. In perturbation
theory this expresses itself in the second order corrections
%--------------------------------------------------------
\begin{eqnarray}
   E_i^{(2)}&=& \sum_{i \neq j} \frac{|\bra{\phi_i^{(0)}}\op{H}'
\ket{\phi_j^{(0)}}|^2}{E_i^{(0)}-E_j^{(0)}}
\ , \label{eq:korrektur2_E}
\end{eqnarray}
%--------------------------------------------------------
which decrease with increasing energetic distance
$E_i^{(0)}-E_j^{(0)}$. In our approximate diagonalization
the diagonal 
%--------------------------------------------------------
\begin{eqnarray}
  \bra{\phi_i^{(0)}}\op{H}\ket{\phi_i^{(0)}} &=&  E_i^{(0)} +
  \lambda E_i^{(1)}
\\
   E_i^{(1)}&=&\bra{\phi_i^{(0)}}\op{H}'\ket{\phi_i^{(0)}}
\end{eqnarray}
%--------------------------------------------------------
and off-diagonal terms
$\bra{\phi_i^{(0)}}\op{H}'\ket{\phi_j^{(0)}}$ of perturbation
theory appear as diagonal and off-diagonal matrix elements of
the reduced Hamilton matrix. Therefore, the approximate
diagonalization includes zeroth and first order by definition and
all higher orders partially up to the cutoff. The inclusion of
eigenstates belonging to degenerate eigenvalues of $\op{H}_0$
poses no problem in our scheme. One should only include all
eigenstates of a degenerate eigenvalue into the approximate
diagonalization, otherwise the convergence is unnecessarily
deteriorated.

%%%%%%%%%%%%%%%%%%%%%%%%%%%%%%%%%%%%%%%%%%%%%%%%%%%%%%%%%%%%%%%%%%%%%%%%
\subsection{Approximate diagonalization based on the rotational-band model}

As a zeroth-order approximation $\op{H}_0$ of the isotropic
Heisenberg Hamiltonian \fmref{eq:Heisenberg_WW} the
rotational-band Hamiltonian\cite{ScL:PRB00,Wal:PRB01,SLM:EPL01}
%--------------------------------------------------------
\begin{equation} \label{eq:H_rb_allgemein}
   \op{H}_0\equiv \op{H}_\text{RB} = -\frac{D J}{2N} \left[\vecop{S}^2 - \sum_
     {n=1}^{N_s} \vecop{S}^2_n \right]
\ .
\end{equation}
%--------------------------------------------------------
is chosen.\cite{Wal:PRB07} This choice rests on the observation
that in bipartite antiferromagnetic spin systems the lowest
eigenvalues within subspaces of total spin $S$ follow the
\textit{Land\'{e}-rule},\cite{LGC:PRB97A,LGC:PRB97B} i.e.
%--------------------------------------------------------
\begin{equation} \label{eq:lande}
   E_\text{min}(S) - E_0 \propto S(S+1) \ .
\end{equation}
%--------------------------------------------------------
The prefactor $-\frac{D J}{2N}$ in Eq. (\ref{eq:H_rb_allgemein})
can be seen as an effective exchange constant which couples the
sublattice spins $\vecop{S}_n$ to the total spin $\vecop{S}$ of
the system. The value of $D$ 
%--------------------------------------------------------
\begin{equation}
   D = 2 \cdot \frac{N_b}{N} \cdot \frac{1}{1-1/N_s}
\end{equation}
%--------------------------------------------------------
is chosen to match the energy of the ferromagnetic state of the
system described by an isotropic Heisenberg
Hamiltonian.\cite{ScL:PRB00} $N_s$ denotes the number of
sublattices which the classical ground state of the system is
composed of, $N_b$ represents the number of bonds of the
system. The eigenstates of $\op{H}_\text{RB}$ are analytically
given in the form $\ket{S_1 \, \dots \, S_{N_s} \, S \, M}$,
which is an enormous advantage for the following
calculations. The corresponding eigenvalues are
%--------------------------------------------------------
\begin{eqnarray} \label{eq:E_RB}
   && E_\text{RB}(S_1,\dots,S_{N_s},S) =\nonumber \\
   && \quad - \frac{D J}{2 N} \left[ S(S+1) 
- \sum_{n=1}^{N_s} S_n \left( S_n + 1 \right) \right]
\ .
\end{eqnarray}
%--------------------------------------------------------
The spectrum of the rotational-band Hamiltonian consists of
eigenvalues that form parabolas, so-called rotational-bands. In
the following a rotational-band is defined as a set of
eventually energetically degenerate eigenstates $\ket{S_1 \,
\dots S_{N_s} \, S \, M}$ with fixed values of quantum numbers
$S_n$ of the sublattice spins.

%===================    figure   =================================
\begin{figure}[ht]
   \centering
   \includegraphics[width=7.5cm]{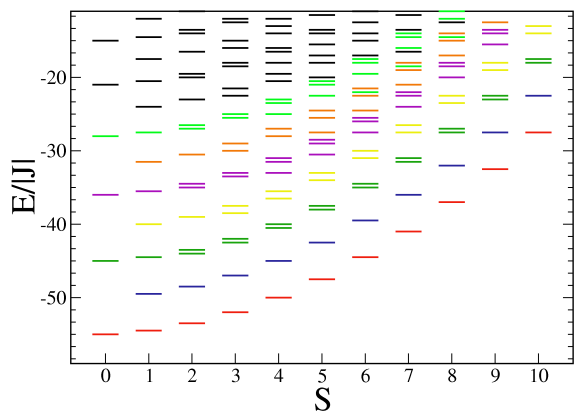}
   \caption{(Color online) Part of the energy spectra of the rotational-band
   Hamiltonian for a antiferromagnetic spin ring $N=8$,
   $s=5/2$. Seven super-bands are colored.}
   \label{F-6}
\end{figure}
%===================    figure   =================================

Figure~\ref{F-6} shows the spectrum of the
rotational-band Hamiltonian for a spin ring of $N=8$ spins with
$s=5/2$. The lowest bands refer to a sublattice spin
configuration of $S_1=S_2=4 \cdot 5/2 = 10$. The next bands
result from a deviation of one sublattice spin from its maximum
value $S_{n,\text{max}}=N/N_s \cdot s$. In such a way the whole
spectrum can be constructed following Eq. \fmref{eq:E_RB}. The
eigenstates of the rotational-band Hamiltonian are highly
degenerate due to the many possibilities of combining single
spins $\vecop{s}_i$ to the sublattice spins $\vecop{S}_n$ and
further on sublattice spins $\vecop{S}_n$ to the total spin
$\vecop{S}$. 

Figure~\ref{F-6} also shows that the rotational-band
spectrum is clustered into \textit{super-bands} (highlighted by
color). A super-band contains those rotational-bands for which
the sum of sublattices spin quantum numbers is the same. One
clearly sees that within the rotational-band spectrum the
low-lying super-bands are well separated.

Inserting $\op{H}_\text{RB}$ into Eq. \fmref{eq:approxDiag_gen}
yields
%--------------------------------------------------------
\begin{equation}
   \op{H}=\op{H}_\text{Heisenberg} = \op{H}_\text{RB} + \op{H}'
\end{equation}
%--------------------------------------------------------
as a starting point for an approximate diagonalization. With
respect to computational resources and due to the fact that the
eigenstates of $\op{H}_\text{RB}$ are given in the form
$\ket{S_1 \, \dots \, S_{N_s} \, S \, M}$ the diagonalization is
performed in subspaces $\mathcal{H}(S,M=S)$ using the
irreducible tensor operator technique. In addition, point-group
symmetries can be used for a further reduction of the
dimensionality. However, only those point-groups can be applied
which do not alter the sublattice structure, i.e. do not lead to
rotational-bands that are not included in the approximate basis
set. Then the symmetry operations on a state belonging to a
certain rotational-band will always produce states which belong
to the same band.

%%%%%%%%%%%%%%%%%%%%%%%%%%%%%%%%%%%%%%%%%%%%%%%%%%%%%%%%%%%%%%%%%%%%%%%%
\section{Bipartite systems - spin ring}
\label{sec-5}

%%%%%%%%%%%%%%%%%%%%%%%%%%%%%%%%%%%%%%%%%%%%%%%%%%%%%%%%%%%%%%%%%%%%%%%%
\subsection{Convergence}

%===================    figure   =================================
\begin{figure}[ht]
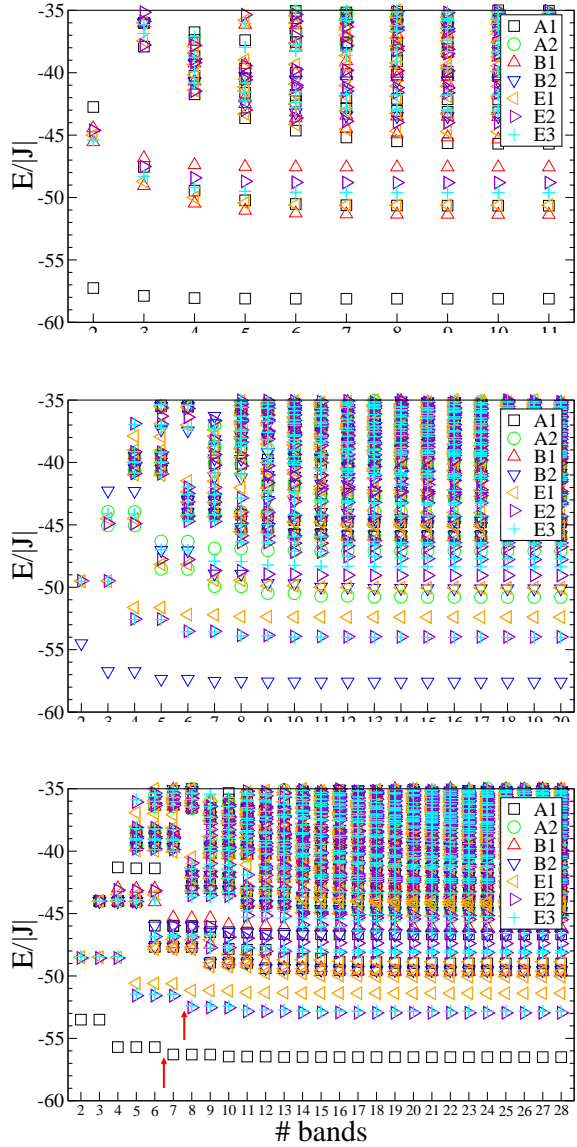
 
   \centering
   \includegraphics[width=7.5cm]{ito-fig-7a.eps} \\[0.3cm] %\hspace{0.3cm}
   \includegraphics[width=7.5cm]{ito-fig-7b.eps} \\[0.3cm] %\hspace{0.3cm}
   \includegraphics[width=7.5cm]{ito-fig-7c.eps}
   \caption{(Color online) Energy levels of an a antiferromagnetic spin ring
   $N=8$ with $s=5/2$ as a function of the number of occupied
   rotational-bands used for diagonalization in subspaces $S=0$
   (top), $S=1$ (center) and $S=2$ (bottom). The arrows in the
   $S=2$ subspace refer to the steps within the convergence
   behavior mentioned in the text. The states are labeled according to irreducible representations of $D_8$.} 
   \label{F-7}
\end{figure}
%===================    figure   =================================

In the following we discuss the properties of the proposed
approximate diagonalization for the example of an
antiferromagnetic spin ring of $N=8$ spins with
$s=5/2$. Figure~\ref{F-7} shows the convergence of
the energy levels. In order to label the levels the full
symmetry group $D_8$ of an octagon was used. One clearly sees
that the convergence within the $S=0$ subspace is fast and
smooth (looking almost exponential).

In subspaces of $S=1$ and $S=2$ the convergence is also fast,
but when only few bands are incorporated sharp steps can be
observed. This is highlighted by two arrows in the bottom graph
of \figref{F-7}. The stepwise convergence is
continued in subspaces with $S > 2$ in a very regular way. It
can be observed that with increasing energy within a certain
subspace $\mathcal{H}(S,M=S)$ the steps are slightly washed
out. The occurrence of the steps depends on the rotational-band
the states are belonging to. For example, the energy of the
lowest state (i.e. the first rotational-band) within
$\mathcal{H}(S=2,M=2)$ is decreasing if 7 bands are incorporated
into the approximate diagonalization while the energies of
states belonging to the second rotational-band are lowered if 8
bands are incorporated, see also discussion in
Sec.~\ref{sec-5-b}. 

%===================    figure   =================================
\begin{figure}[ht]
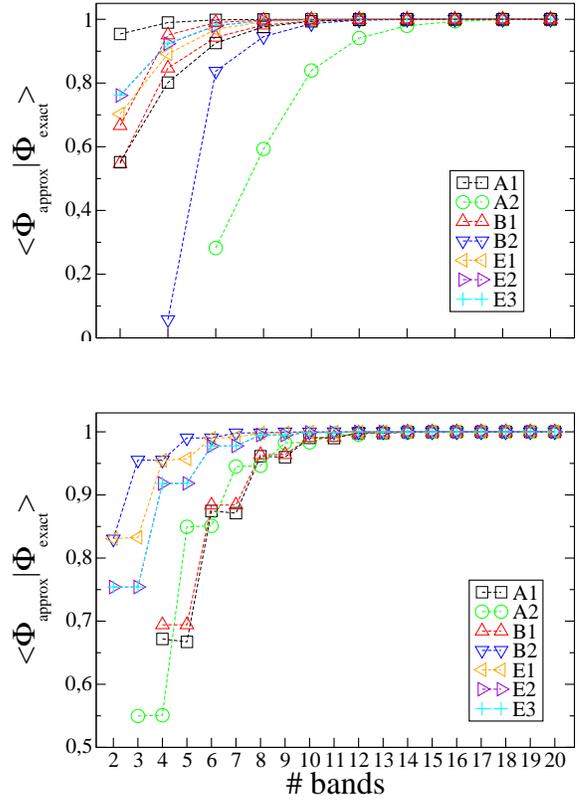
 
   \centering
   \includegraphics[width=7.5cm]{ito-fig-8a.eps} \\[0.3cm] %\hspace{0.3cm}
   \includegraphics[width=7.5cm]{ito-fig-8b.eps}
   \caption{(Color online) Convergence of the eigenstates of an a
   antiferromagnetic spin ring $N=8$ with $s=5/2$ as a function
   of the number of occupied rotational-bands used for
   diagonalization in subspaces $S=0$ (top) and $S=1$ (bottom). The states are labeled according to irreducible representations of $D_8$.}
   \label{F-8}
\end{figure}
%===================    figure   =================================

In Fig. \ref{F-8} the convergence of some
low-lying eigenstates of this spin ring are presented. The
convergence behaves in analogy to the convergence of the
eigenvalues. The stepwise convergence in $S=1$ becomes
obvious. Nevertheless, while using only a fraction of basis
states (approximately $30\%$ of the basis states within each
subspace) the approximate low-lying eigenstates are practically
converged against the exact eigenstates. In addition, it can be
seen that states of higher energy converge slower than
low-lying states.

%===================    figure   =================================
\begin{figure}[ht] 
   \centering
   \includegraphics[width=7.5cm]{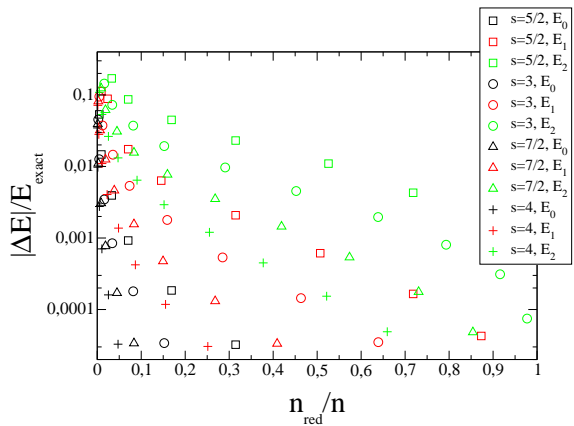}
   \caption{(Color online) Relative difference between
   approximate and exact energy eigenvalues for the lowest
   states of the first three occupied rotational-bands for an
   antiferromagnetic spin ring $N=8$ with various $s$ and total
   spin $S=0$. $E_0$ refers to the energy of the lowest
   state of the first rotational-band, $E_1$ to the lowest state
   of the second rotational-band and $E_3$ to the lowest state
   of the third rotational-band respectively.}
   \label{F-9}
\end{figure}
%===================    figure   =================================

We also investigate the convergence for various single spin
quantum number $s$. In Fig. \ref{F-9} the relative
difference between the approximate energy values and the exact
values is displayed for various $s$ in the subspace $S=0$. The
levels which have been chosen belong to the first three occupied
rotational-bands. One clearly sees that the approximate
diagonalization converges more rapidly the higher the single
spin is. This is not surprising since the rotational-band model
\fmref{eq:H_rb_allgemein}, which is based on classical
assumptions, is itself more accurate the larger $s$ is.

%%%%%%%%%%%%%%%%%%%%%%%%%%%%%%%%%%%%%%%%%%%%%%%%%%%%%%%%%%%%%%%%%%%%%%%%
\subsection{Approximate selection rule}
\label{sec-5-b}

It turns out that the aforementioned occurrence of steps can be
understood and even be employed for a further reduction of the
size of Hamilton matrices. The underlying reason is that the
full Hamiltonian connects states belonging to different
rotational-bands with very different strength. After having
inspected the reduced Hamilton matrices of various bipartite
systems we arrive at the following empirical selection rule.

The matrix elements $\bra{S_{1,a} \, S_{2,a} \, S \, M}\op{H}\ket{S_{1,b}
  \, S_{2,b} \, S \, M}$ of the full Hamiltonian between
  rotational-band states are (several) orders of magnitude
  bigger than all other matrix elements if
%--------------------------------------------------------
\begin{equation} \label{eq:selection_ring}
   |S_{1,a} - S_{2,a}| - |S_{1,b} - S_{2,b}| = 0 \ .
\end{equation}
%--------------------------------------------------------
Here $S_{1,a}$ and $S_{2,a}$ denote the total spins of
sublattices one and two in $\bra{S_{1,a} \, S_{2,a} \, S \, M}$,
respectively. Matrix elements that are not compatible with this
rule can be neglected which (after a proper rearrangement)
results in a new block-diagonal structure of the reduced
Hamilton matrix. These blocks are of smaller size and can be
diagonalized separately.

%%%%%%%%%%%%%%%%%%%%%%%%%%%%%%%%%%%%%%%%%%%%%%%%%%%%%%%%%%%%%%%%%%%%%%%%
\subsection{Application to $\{\text{Fe}_{12}\}$}
\label{sec-5-c}

We now apply the approximate diagonalization to an existing
molecular spin ring\cite{CCF:ACIE99} that contains 12
$\text{Fe}^{3+}$ ions with $s=5/2$. The system can be modeled by
an isotropic Heisenberg Hamiltonian with antiferromagnetic
nearest-neighbor coupling $J$.\cite{CCF:ACIE99,IAA:JPSJ03} It
was theoretically investigated in Ref.~\onlinecite{EnL:PRB06}
with the help of QMC methods; the exchange parameter was
determined to be $J=35.2~\text{K}$.

Our intention is to show that it is advantageous to combine a
stochastic method such as QMC and an exact or approximate
diagonalization. In such a combination the role of QMC would be
to determine the exchange parameters from thermodynamical
observables as done in Ref.~\onlinecite{EnL:PRB06}. For large
systems this is practically impossible using exact or
approximate diagonalization since diagonalization requires an
enormous numerical effort whereas QMC methods scale much more
favorable with system size for bipartite systems or even
frustrated systems above a certain temperature. The role of
exact or approximate diagonalization then would be to use the
exchange parameters obtained by QMC for the evaluation of the
energy spectrum which then can be used e.g. to interpret INS
measurements.\cite{BBC:CPC03, BCC:IC99}

%===================    figure   =================================
\begin{figure}[ht] 
   \centering
   \includegraphics[width=7.5cm]{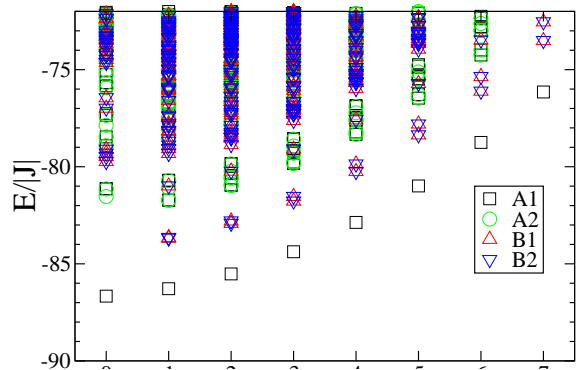} \\[0.3cm]
   \includegraphics[width=7.5cm]{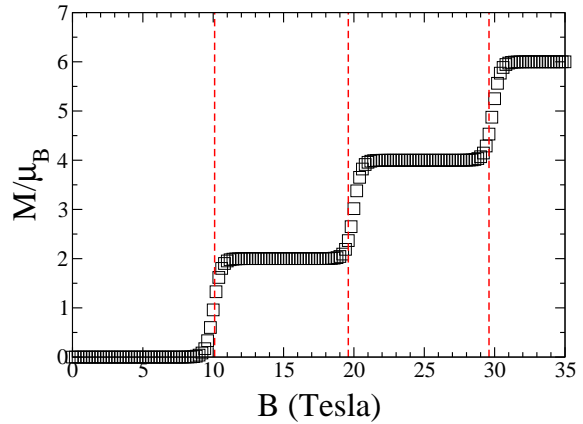}
   \caption{(Color online) Approximate spectrum of a spin ring with $N=12$ spins
   $s=5/2$ calculated using 8 bands, $D_2$ point-group
   symmetry and the approximate selection rule in
   Eq. (\ref{eq:selection_ring}) (top).  Corresponding
   magnetization of the system (bottom). The dashed red lines
   refer to experimental data of the first three magnetization
   steps from Ref.~\onlinecite{IAA:JPSJ03} with
   $J=35.2~\text{K}$ and $k_B T/J=0.01$.} 
   \label{F-10}
\end{figure}
%===================    figure   =================================

Figure~\xref{F-10} shows the low-energy part of the
approximate spectrum of the $\{\text{Fe}_{12}\}$ compound
modeled by an isotropic Heisenberg Hamiltonian. For the
approximate calculation of the spectrum full rotational symmetry
as well as $D_2$ point-group symmetry are used. The calculations
are performed using 8 occupied rotational-bands in the $S=0$
subspace and the corresponding number of bands in subspaces with
$S>0$. Overall $21,570,976$ states have been taken into account,
which are only about 15~\% of all basis states
($\text{dim}(\mathcal{H})=144,840,476$). Additionally the
approximate selection rule given in
Eq. \fmref{eq:selection_ring} was used in order to reduce the
dimensionality of the matrices which have to be diagonalized.
Figure~\xref{F-10} also displays the magnetization curve
which of course can be obtained for a bipartite system by QMC as
well. The magnetization steps\cite{IAA:JPSJ03} can be reproduced
using the approximate diagonalization.

%%%%%%%%%%%%%%%%%%%%%%%%%%%%%%%%%%%%%%%%%%%%%%%%%%%%%%%%%%%%%%%%%%%%%%%%
\subsection{Next-nearest-neighbor coupling -- introducing frustration}

In the previous parts we demonstrate that the approximate
diagonalization scheme based on the rotational-band Hamiltonian
yields good results for bipartite, i.e. unfrustrated
antiferromagnetic spin systems. We now want to investigate how
robust the approximate diagonalization is against the
introduction of frustration. To this end we study a spin ring
with $N=8$ and $s=5/2$ with antiferromagnetic nearest-neighbor
coupling $J=J_{nn}$ and an additional antiferromagnetic
next-nearest-neighbor-coupling $J_{nnn}$ which acts
frustrating. In a corresponding classical system the N\'{e}el
state (up-down-up-down- \dots) would no longer be the ground
state, instead canting can occur. One can qualitatively say that
with increasing  $J_{nnn}/J_{nn}$ also the frustration
increases. 

%===================    figure   =================================
\begin{figure}[ht]
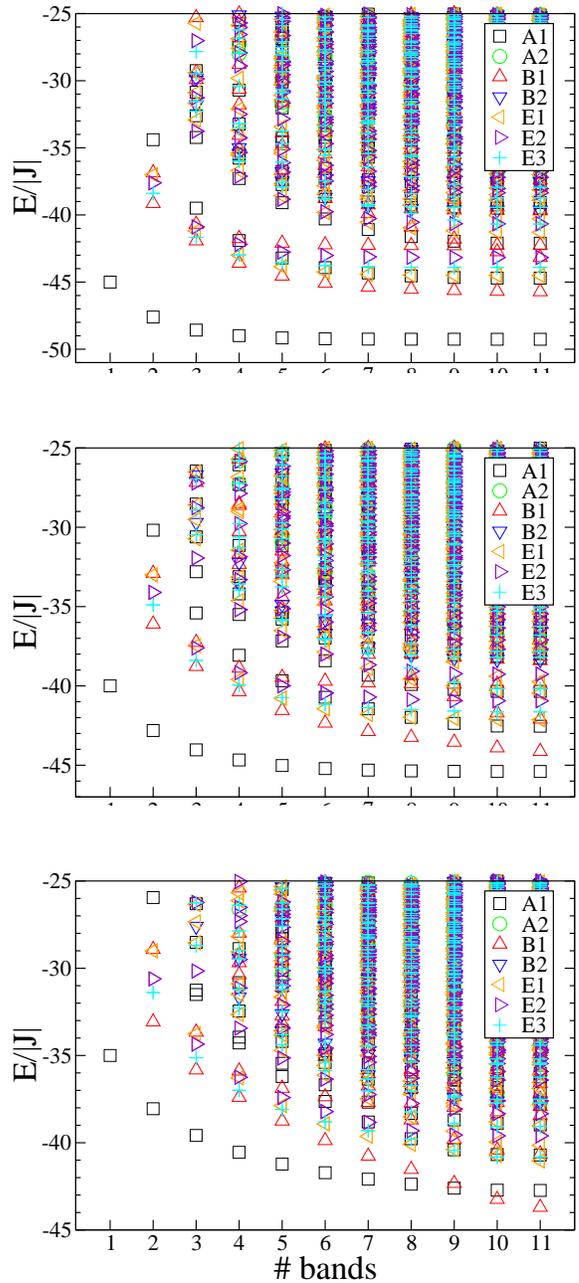
 
   \centering
   \includegraphics[width=7.5cm]{ito-fig-11a.eps} \\[0.3cm] %\hspace{0.3cm}
   \includegraphics[width=7.5cm]{ito-fig-11b.eps} \\[0.3cm] %\hspace{0.3cm}
   \includegraphics[width=7.5cm]{ito-fig-11c.eps}
   \caption{(Color online) Energy levels of an a antiferromagnetic spin ring
   $N=8$ with $s=5/2$ and additional next-nearest-neighbor
   coupling $J_{nnn}$ as a function of the number of occupied
   rotational-bands used for diagonalization in subspaces $S=0$
   with $J_{nnn}/J_{nn}=0.2$ (top), $J_{nnn}/J_{nn}=0.3$
   (center) and $J_{nnn}/J_{nn}=0.4$ (bottom). The states are labeled according to irreducible representations of $D_8$.}
   \label{F-11}
\end{figure}
%===================    figure   =================================

Figure~\ref{F-11} displays the effect of $J_{nnn}$ in the
subspace $\mathcal{H}(S=0,M=S)$ for the same system that is
discussed in \figref{F-7} for $J_{nnn}=0$. The energy gap
between the ground state and the first excited state decreases
with increasing frustration. Moreover, the convergence of the
ground state as well as of excited states becomes slower. With
$J_{nnn}/J_{nn}=0.4$ the convergence is rather poor and the
quantum mechanical ground state now belongs to the irreducible
representation $B_1$ of the symmetry group $D_8$. This means
that the true ground state is not the result of an adiabatic
continuation ($\lambda:0\rightarrow 1$ in
Eq.~\fmref{eq:approxDiag_gen}) from the ground state of the
rotational-band model, which belongs to $A_1$. We just like to
mention for the interested reader, that this change of the
character of the ground state constitutes a so-called Quantum
Phase Transition; in this case for the antiferromagnetic chain
with next-nearest-neighbor exchange.

Summarizing, if frustration is only small the approximate
diagonalization still yields good results. Moreover, the
approximate selection rule \fmref{eq:selection_ring} is also
applicable which is very helpful in calculating the full
spectrum of the system.

%%%%%%%%%%%%%%%%%%%%%%%%%%%%%%%%%%%%%%%%%%%%%%%%%%%%%%%%%%%%%%%%%%%%%%%%
\section{Summary}

In this work we have demonstrated that the full spin-rotational
symmetry can be combined with arbitrary point-group
symmetries. This enables us to obtain exactly the complete
energy spectrum of Heisenberg spin systems for so far
unprecedented system sizes. Moreover, we have outlined a scheme
to approximately diagonalize the Hamilton matrix again using 
the full spin-rotational symmetry and point-group
symmetries. This approximation works well for bipartite
antiferromagnetic spin systems. For frustrated systems the
quality reduces with increasing frustration. How such a scheme
can be refined for frustrated systems will be the subject of
future investigations.

%%%%%%%%%%%%%%%%%%%%%%%%%%%%%%%%%%%%%%%%%%%%%%%%%%%%%%%%%%%%%%%%%%%%%%%%
\acknowledgments{Computing time at the Leibniz Computing Center
in Garching is gratefully acknowledged as well as helpful
advices of Dieter an Mey and Christian Terboven of High
Performance Computing Center, RWTH Aachen, in setting up openMP
directives. We also thank Boris Tsukerblat for fruitful
discussions about the irreducible tensor operator technique and
Nedko Ivanov for drawing our attention to some literature. This
work was supported within a Ph.D. program of the State of Lower
Saxony in Osnabr\"uck.}

%\bibliography{js-own,js-mag,rs-mag,rs-ito,rs-swt,rs-own}
%\bibliography{/home/schnack/tex/bibtex/js-own,/home/schnack/tex/bibtex/js-mag,/home/schnack/tex/bibtex/js-mis}

\end{document}